\documentstyle[prd,aps,preprint]{revtex}
\begin{document}
\draft
%
%  Uncomment following two lines and one below for 2 column format.
%@twocolumnfalse\endcsname
\preprint{Nisho-98/2} \title{Radiations from Oscillating Axionic Boson Stars
in an External Magnetic Field} 
\author{Aiichi Iwazaki}
\address{Department of Physics, Nishogakusha University, Shonan Ohi Chiba
  277,\ Japan.} \date{March 29, 1998} \maketitle
\begin{abstract}
We solve numerically a field equation of axions coupled with gravity 
and show solutions representing oscillating axionic boson stars 
with small mass $\sim 10^{-12}M_{\odot}$ and large radius $\sim 10^{8}$ cm.
We present explicitly a relation between the mass and the radius of 
the boson stars with such a small mass.
These axionic boson stars are shown to possess oscillating electric currents
in an external magnetic field and to radiate electromagnetic fields with 
a frequency given by mass of the axions. 
We estimate roughly the luminosity of the radiation in a strong magnetic field 
$10^{12}$G of neutron stars,  it is $\sim 10^{34}$ erg/s with
the axion mass $10^{-5}$eV. 
\end{abstract}
%\pacs{73.61.-r,73.20.Dx,73.40.Hm,73.40.Gk}
%\pacs{14.80.Mz, 98.80.Cq, 95.30.+d, 05.30.Jp, 98.70.-f \\Axion, 
%Boson Star, Dark Matter, Radiation 
%\hspace*{3cm}}
\vskip2pc
%%%%%%%%%%%%%%%%%%%%%%%%%%%%%%%%%%%%%%%%%%%%%%%%
The axion is 
a Goldstone boson associated with 
Peccei-Quinn symmetry\cite{PQ} and is one of most plausible candidates of 
the dark matter in the Universe. 
The axions are produced \cite{kim,text} in early Universe 
mainly by the decay of the axion strings or 
coherent oscillation of the axion field. 
These axions form incoherent axion gas in the present Universe. 
It has been recently argued\cite{kolb} 
that not only the incoherent axion gas but also coherent axionic 
boson stars\cite{re} are produced 
by the coherent oscillation of the axion field.
Namely, inhomogeneity of the coherent oscillation on the scale beyond 
the horizon 
generates axion clumps as well as axion domain walls. The domain walls decay 
soon after their productions and leave a primordial magnetic field\cite{iwaza} 
as well as incoherent axions. 
On the other hand, the axion clumps contract gravitationally 
in later stage of 
the Universe to the axionic boson stars.
We call them simply as axion stars.  

In this letter we present numerical solutions \cite{real}
of the axion star by solving 
a field equation of the axion together with Einstein equations. Our concern
is confirming the existence of the axion stars with quite small masses 
and finding the explicit relation
between the radius ( or the typical value of the axion field ) 
and the mass of the star. This is because 
the mass of the axion star produced by the mechanism mentioned above 
has been estimated to be order of 
$10^{-12}M_{\odot}$ by Kolb and Tkachev\cite{kolb}.
Furthermore, using these solutions, 
we show that the axion stars in an external magnetic field 
can emit electromagnetic radiations with 
a frequency given by the mass of the axion.    
The luminosity of the radiations is roughly $10^{34}$ erg/s 
in the magnetic field $\sim 10^{12}$ G of neutron stars.
  
Now we show solutions of the axion stars 
by analyzing the field equations,

\begin{eqnarray}
\label{a}
\ddot{a}&=&\frac{(\dot{h_t}-\dot{h_r})\dot{a}}{2}+a''
+(\frac{2}{r}+\frac{h_t'-h_r'}{2})a'-m^2a\quad,\\
\label{h_t}
h_t'&=&\frac{h_r}{r}+4\pi Gr(a'^2-m^2a^2+\dot{a}^2)\quad,\\
\label{h_r}
h_r'&=&-\frac{h_r}{r}+4\pi Gr(a'^2+m^2a^2+\dot{a}^2)\quad,
\end{eqnarray}
where we have assumed gravity being small, i.e. $h_{t,r}\ll 1$ 
so that the metric is such that 
$ds^2=(1+h_t)dt^2-(1+h_r)dr^2-r^2(d\theta^2+\sin^2\theta d\phi^2)$;
$r,\theta$, and $\phi$ denote the polar coordinates.
The first equation is the equation of the axion field $a$ 
coupled only with gravity.
The second and the third equations are Einstein equations.  
A dot ( dash ) indicates a derivative in time $t$ ( $r$ ). $G$ ( $m$ ) 
is the graviational constant ( the mass of the axion ).
We have neglected the potential term of the axion
because an amplitude of the field $a$ is assumed to be sufficiently small for 
the nonlinearity of $a$ not to arise. This implies that the mass of the axion 
star is small enough. Actually the masses we are concerned with are 
such as $\sim 10^{-12}M_{\odot}$. We need to 
impose a boundary condition such as 
$h_r(r=0)=0$ for the regularity of the space-time.

Changing the scales such that $\tau=mt$, $x=mr$ and $b=a/m$, we rewrite 
the equations as follows,

\begin{eqnarray}
\label{bd}
\ddot{b}&=&\dot{V}\dot{b}+b''+(\frac{2}{x}+V')b'-b\quad,\\
\label{V}
V'&\equiv&\frac{h_t'-h_r'}{2}=\epsilon
(\frac{\int_0^xdxx^2(b'^2+\dot{b}^2+b^2)}{x^2}
-xb^2)
\end{eqnarray}
with $\epsilon=4\pi Gm^2$,
where we have expressed $V'$  
in terms of the field $b$, solving eq(\ref{h_t}) and eq(\ref{h_r});
here a dot ( dash ) denote a derivative in $\tau$ ( $x$ ).  
We understand that if the gravitational effect is neglected ( $\epsilon=0$ ),
the equation of $b$ is reduced to a Klein-Gordon equation.
Thus the frequency $\omega$ of the field $b$ is modified from the value
of the free field, $1$ ( $m$ in the physical unit ), only by a 
small quantity proportional to the gravitational effect, 
$\epsilon$; $\omega=1-o(\epsilon)$. 
We look for such a solution \cite{real} that 

\begin{equation}
\label{bB}
b=A_0 B(x)\sin\omega\tau+o(\epsilon)\sin3\omega\tau\quad, 
\end{equation}
where $B(x)$ represents coherent axions 
bounded gravitationally with its spatial extension representing 
the radius of the axion star; $B(x)$ is normalized 
such as $B(x=0)=1$. Later we find that 
$A_0$ is a free parameter determining
a mass or a radius of the axion star.  
The second term is a small correction of the order of
$\epsilon$.
Inserting the formula eq(\ref{bB}) into eq(\ref{bd}) and eq(\ref{V}) and 
taking account of the gravitational effects only with the order of $\epsilon$, 
we find that    

\begin{equation}
\label{B}
k^2B=B''+(\frac{2}{x}+\epsilon A_0^2(T+\frac{3U'}{4}))B'+
\frac{\epsilon A_0^2(U+v)B}{2}
\end{equation}
with 
\begin{equation}
T\equiv\frac{\int_0^xz^2B^2dz}{x^2}\quad \mbox{and}\quad
U\equiv\int_0^xdy(\frac{\int_0^yz^2B'^2dz}{y^2}-yB^2)\quad,
\end{equation}
where $k^2$ ( $=1-\omega^2$ ) is a binding energy of axions. We have imposed a 
boundary condition for the consistency such that 
$V(x=0)=h_t(x=0)/2=\epsilon A_0^2v\omega \sin2\omega\tau$ ; this is 
the definition of constant $v$ in the above formula.

We can see that the parameter $\epsilon A_0^2$ can take an arbitrary value
and that it represents the gravitational effect of this system.
Namely the mass of the axionic boson star is determined by choosing 
a value of the parameter. Note that the normalization of $B$ has been 
fixed in eq(\ref{B}) although the equation is a linear in $B$.

We may take the value of $v$ without loss of generality such that 
$v=-U(x=\infty)$. Then the inverse $k^{-1}$ of the binding energy 
is turned out to represent
a radius of the axion star; $B$ decays exponentially 
such as $\exp(-kx)$ for $x\to\infty$.  
It turns out from eq(\ref{B}) that 
the choice of small values of $\epsilon A_0^2$ lead to solutions representing 
the axion stars with small mass and large radius $k^{-1}$.

Before solving eq(\ref{B}) numerically, it is interesting to rewrite the 
equation as following. That is, we rewrite the equation by 
taking only dominant terms of a ``potential'', $V_b$, in eq(\ref{B}),

\begin{equation}
\label{Vb}
V_b=\epsilon A_0^2((T+\frac{3U'}{4})B'+\frac{(U+v)B}{2})
\end{equation}
in the limit of 
the large length scale; setting $x=\lambda y$, we take a dominant term  
as $\lambda \to \infty$. 
This corresponds to thinking the axion stars 
with their spatial extension being large. Note that our concerns are 
such axion stars with small mass and 
with large radius.

Then, since the dominant term in the limit is 
the last term in eq(\ref{Vb}), 
$U+v\sim \int_x^{\infty}dxxB^2$,
we obtain the following equation,

\begin{equation}
\label{BB}
\bar{B}=\bar{B}''+\frac{2\bar{B}'}{z}+
\frac{\bar{B}\int_z^{\infty}dyy\bar{B}^2}{2}\quad,
\end{equation}
where we have scaled the variables such that 
$B^2=k^2\bar{B}^2/\epsilon A_0^2$ and $x=k^{-1}z$;
a dash denotes a derivative in $z$.
This equation is much simpler than eq(\ref{B}), where we need to find each 
eigenvalue of $k$ for each value of $\epsilon A_0^2$ given, in order to obtain 
solutions of the axion stars with various masses. On the other hand,
we need only to find an appropriate value of 
$\bar{B}(z=0)=k^2/\epsilon A_0^2$ in order to obtain such solutions.
A relevant solution we need to find is the solution without any nodes.
Obviously, the solution is characterized by 
one free parameter, $k^2$ or $\epsilon A_0^2$, which is related to the mass 
of the axion star. Namely the choice of a value of the mass determines 
uniquely the properties of the axion star,
e.g. radius of the star, distribution of 
axion field $a$, e.t.c.. 

Although the equation (\ref{BB}) 
governs the axion star only with the large radius, 
we have confirmed by solving numerically original equation (\ref{B}) that 
the stars of our concern,
whose masses are at most $10^{-12}M_{\odot}$, can be controlled 
by the equation (\ref{BB}). Thus we may 
determine the explicit relation between the mass 
and the radius of the axion star, by finding 
the above solution of eq(\ref{BB}),
   
\begin{equation}
\label{MR}
M=6.4\,\frac{m_{pl}^2}{m^2R}\quad,
\end{equation}
where $m_{pl}$ is Planck mass.
Numerically we can see that for example, 
$R=1.6\times10^5m_5^{-2}\mbox{cm}$ 
for $M=10^{-9}M_{\odot}$, 
$R=1.6\times10^8m_5^{-2}\mbox{cm}$ for $M=10^{-12}M_{\odot}$,
e.t.c. with $m_5\equiv m/10^{-5}\mbox{eV}$.
We have also numerically confirmed that the axion field $a$ of the 
solution can be 
practically parameterized such that $a=fa_0\exp(-r/R)$ where $f$ is the decay 
constant of the axion whose value is constrained from cosmological
and astrophysical considerations\cite{text}; 
$10^{10}$ GeV $\le$ $f$ $\le$ $10^{12}$ GeV. This implies a constraint 
on the mass of the axion; $10^{-5}\mbox{eV}<m<10^{-3}\mbox{eV}$.
Amplitude $a_0$ ( $=mA_0/f$ ) is found to be expressed 
in terms of the radius $R$,

\begin{equation}
\label{aR}
a_0=1.73\times 10^{-8}\frac{(10^8\mbox{cm})^2}{R^2}\,
\frac{10^{-5}\mbox{eV}}{m}\quad.
\end{equation}  

In this way we have found solutions representing the axion stars with their 
mass $M$ being typically $10^{-12}M_{\odot}$ 
and with the radius $R$ being $10^{8}$cm.
The binding energy is turned out to be extremely small so that 
the frequency $\omega$ of the field $a$ is given by $m$.

Next, we wish to show that 
the axion stars can emit electromagnetic radiations
with energy $m$ in an external magnetic field. The reason is that 
the coherent axion field gain an oscillating current in the magnetic field
owing to the following interaction,

\begin{equation}
   L_{a\gamma\gamma}=\frac{c\alpha a\vec{E}\cdot\vec{B}}{f\pi}
\label{EB}
\end{equation}
with $\alpha=1/137$,
where $\vec{E}$ and $\vec{B}$ denote electric and 
magnetic field, respectively. 
The value of $c$ depends on the axion models\cite{DFSZ,hadron};
typically it is the order of one.

It is easy to see from this interaction that the coherent axion
field may have electric charge density 
$-c\alpha\vec{\partial}a\cdot\vec{B}/f\pi$ 
in the magnetic field $\vec{B}$\cite{Si}. 
We assume that the field $\vec{B}=(0,0,B)$ is spatially uniform and 
that the geometry of the axion field $a$  
representing the boson stars is spherical. Then we understand easily that 
this axion field has a charge distribution such that 
it has negative charges on a hemisphere 
( $z>0$ ) and positive charges on the other side of the sphere ( $z<0$ ). 
Net charge is zero. Thus, the star possesses the electric field, $E$,
parallel to the magnetic field associated with 
the charge distribution. As we have shown before, the field $a$
representing the axion star oscillates with frequency being approximately
given by $m$. Therefore, the charge distribution oscillates with the frequency
and so an electric current associated with the charge oscillates similarly.
This fact induces emission of the electromagnetic fields. 

The radiation from the axion stars can be understood in another way.
Since a mixing angle\cite{mix} of an axion and a photon generated 
by the interaction eq(\ref{EB}) 
is the order of $c\alpha B/fm\pi$ in the magnetic field $B$, 
equations of motion of gauge potentials $A_i$ can be derived such as 
$\partial^2 A_i=\partial^2 c\alpha B_ia/fm\pi \sim c\alpha B_im a/f\pi$. 
Hence, the oscillation of this current,
$c\alpha B_im a/f\pi$, produces the radiation as 
dictated above.

It is easy to evaluate the luminosity $L$ of the radiation by assuming 
the configuration of the field $a$ such that $a=fa_0\exp(-r/R)$ whose
form we have confirmed numerically. 
Noting the wave length ( $=m^{-1}$ ) of the radiation is 
much smaller than the radius $R$ of the axion star, we obtain

\begin{equation}
\label{L}
L=\frac{64B^2a_0^2c^2\alpha^2}{3m^4R^2}\quad,
\end{equation} 
where we have assumed 
that whole of the axion star is involved in the external magnetic field.
On the other hand, when the length scale of the magnetized objects 
like neutron stars is smaller than the axion stars, $R$ in the above 
formula eq(\ref{L}) should be understood to denote the radius 
of the magnetized objects. Especially what we are concerned with is 
the collision of a neutron star with the axion star
whose radius is $10^8\mbox{cm}\sim10^{10}\mbox{cm}$. This radius is 
$10^2\sim10^4$ times 
larger than the radius of the neutron star.   
Thus $R$ in eq(\ref{L}) denotes the radius of the neutron star in such an 
example. 

We understand that since the luminosity $L$ is proportional to $B^2$, 
it is larger as the axion star is exposed to a stronger magnetic field.
Furthermore, as $L$ depends on $R^{-2}$, it is larger as the radius of 
the axion star or the magnetized object is smaller. Hence we understand 
that the neutron star among astrophysical objects yields large luminosity 
when it collides with the axion star. We estimate the luminosity of the case, 

\begin{equation}
L=2.7\times 10^{4}c^2\,\mbox{erg/s}\,\frac{B^2}{(10^{12}\mbox{G})^2}\,
\frac{M^4}{(10^{-12}M_{\odot})^4}\,
\frac{(10^6\mbox{cm})^2}{R^2}\quad,
\end{equation}
where we have used the above formulae eq(\ref{MR}) and eq(\ref{aR}) for 
expressing $a_0$ in terms of the mass $M$ of the axion star.
We note that this luminosity is that of the monochromatic radiation with the 
frequency, $m/2\pi=2.4\times 10^{9}$Hz$\,(m/10^{-5}\mbox{eV})$. 
This is much weaker than a corresponding luminosity of synchrotron radiations 
with the frequency from the neutron stars.
This emission continues until the axion star passes through the neutron stars.
It takes 
$10^8$cm $\times(10^{-12}M_{\odot}/M) /(3\times10^7$cm/s) $\sim10\,\mbox{sec}$ 
$(10^{-12}M_{\odot}/M)$,
assuming that the velocity of the axion star is 
$3\times10^7$cm/s, which is the typical 
velocity of matters composing the halo in our galaxy.
Thus it is impossible to observe the radiation. 

Until now we do not take into account 
the influence of the medium of the astrophysical object such 
as the neutron star.
When the medium is electric conducting, we need to consider 
its effects on the radiations or the electric field associated with the axion
star.
Here we would like to mention that the electric field possessed 
by the axion star in an external magnetic field induces 
an electric current in 
an astrophysical medium. This current also oscillates and emits a radiation.
Since the strength of the current is proportional to 
an electric conductivity, $\sigma$, 
of the medium, the medium with large conductivity such as the crust of 
the neutron star leads to strong radiations. 
In the case, the luminosity is approximately given by

\begin{equation}
L_{\sigma}=(\frac{\sigma}{m})^2L
\end{equation} 
where the conductivity should be taken 
as an conductivity averaged over the volume 
of the medium. For instance, in the case of the neutron star  
the conductivity of its crust has been estimated\cite{con}
theoretically such as $\sigma=$O($10^{24}$/s). When we adopt this value
for $\sigma$, the luminosity $L_{\sigma}$ is given by  
$\sim10^{34}$ erg/s. Although this is a rough estimation,
we can expect that the neutron star emit sufficiently strong radiations
to be detectable when it collides with the axion star. The detail analysis
is in progress.

%It is practically useful to rewrite the luminosity as 
%an intensity of the radiation received at the earth,

%\begin{equation}
%I=5.6\times 10^{-7}c^2\,\mbox{Jy}\,\frac{B^2}{(10^{12}\mbox{G})^2}\,
%\frac{M^4}{(10^{-12}M_{\odot})^4}\,
%\frac{m^3}{(10^{-5}\mbox{eV})^3}\,\frac{(1\mbox{kpc})^2}{r^2}
%\end{equation}
%where we have assumed that the neutron star or the white dwarf 
%is located at the distance $r$ from the earth; 
%$1$ Jy $=10^{-23}$ erg cm$^{-2}$ s$^{-1}$ Hz$^{-1}$.
%This intensity is much smaller than 
%those of synchrotron radiations from pulsars.

Finally we point out that the monochromatic radiation 
under the discussion should be  
detected for confirmation of the phenomena associated with the axion
just after the discovery of rebrightness of dark neutron stars
or white dwarfs caused by the axion star. The rebrightness of these stars 
has been discussed in the previous paper\cite{iwa} 
to be caused by energy dissipation 
of the axion star when it collide with them. 
These two phenomena, monochromatic radiations and 
thermal radiations owing to the energy dissipation, are distinctive in 
the collision of the axion star with these strongly magnetized stars.

In conclusion, we have shown the solutions of the oscillating 
axionic boson stars with small masses. We have made explicit 
the relation between the mass and the radius of these axion stars.
They have also been shown to gain two types of oscillating electric currents 
in the magnetized conducting medium; currents made of axions themselves
and currents made of electrons in the medium.
Both of them emit monochromatic radiations with the frequency 
given by the mass of the axion. Having estimated the intensity of the 
radiations, we found that the radiations
by the currents of electrons are strong enough to be detectable.
But, since the duration of the emission is much short, 
it seems difficult to observe the radiations.

\end{document}